# What's Wrong with Net-Promoter Score?

Nicholas I. Fisher[1] and Raymond E. Kordupleski

**ABSTRACT**

Net-Promoter Score (NPS) is now ubiquitous as an easily-collected market research metric, having displaced many serious market research processes.  Unfortunately, this has been its sole success.  It possesses few, if any, of the characteristics that might be regarded as highly desirable in a high-level market research metric; on the contrary, it has done considerable damage both to companies and to their customers.

---

[1] Nicholas Fisher is Visiting Professor of Statistics, School of Mathematics & Statistics, University of Sydney, NSW 2006 AUSTRALIA (email Nicholas.Fisher@sydney.edu.au), and Principal, ValueMetrics Australia; and Raymond Kordupleski was formerly AT&T Customer Satisfaction Director; now retired in Louisville, Colorado, 80027 USA.  The authors thank Stephen Sasse for helpful comments.

**Introduction**

Here is a scenario from everyday life.

You've just got completed a small online transaction. Before you log out, you are presented with a request to answer a one question customer survey:

*On a scale of 0 to 10, where* 0 = Not at all *and* 10 = Definitely*, please rate your willingness to recommend us to others.*

As a statistician, this should trouble you.

This familiar scene is an example of the use of NPS (Net-Promoter Score), a ubiquitous metric collected routinely by companies after most transactions. We contend that statisticians should be taking a stand against the use of NPS as a panacea for the problems it purports to address, including its use as a stand-alone metric in assessing customer satisfaction or staff satisfaction.

To expose NPS's failings, we shall focus primarily on a customer (rather than a staff member) as the stakeholder being surveyed. The customer may be a consumer, or may be a corporation. We will be making a critical distinction two different types of survey respondents: on the one hand, the person who actually makes the purchasing decision about a product or service; and on the other, the person who actually uses it. These may be coincident, or maybe not. Thus, in a corporation, the purchasing decision about an automobile may be made by a senior executive, but the transaction carried out by an administrative assistant. In the consumer world, a purchasing decision about a teenager's first car may be made by a parent (taking account of price, Safety, …) whereas the teenager may be making the actual choice of vehicle and have much of the experience of interacting with the vendor.

Section 2 outlines key goals of customer satisfaction surveys, and provides a brief description of a long-established and proven process to achieve these goals. Section 3 gives a description of NPS and evaluates it as market research metric.

1. **Customer perception survey guidelines**

The customer survey process is an important means of managing a company's relationship with customers. Key attributes of a good survey process include:

   a. a statistically sound method, that provides assurance that no attributes of the product or services that are important to the customer have been omitted from the survey, and results in acquiring reliable data;

   b. a means of linking survey results to higher-level business drivers;

   c. actionable survey results, including the ability to drill down;

   d. a means of identifying where to focus improvement priorities in order to have the greatest beneficial impact on both customers and the business' bottom line;

   e. Comparable and useful benchmarking metrics.

A process to design and conduct customer satisfaction surveys that meets these criteria was developed some 30 years ago at AT&T … in response to a crisis. The crisis is described in detail by Kordupleski (2003, xv *et seq*.), and subsequently summarized in Fisher (2013, Chapter 4) as follows:

> In the mid-1980s, AT&T was confronted by a paradox: on the one hand, customer satisfaction levels were running at about 95%; on the other hand, they lost 6% market share, where 1% was worth $600,000,000. For the first time in corporate history, AT&T laid people off — 25,000

worldwide from an overall staff of 300,000 — including managers recently rewarded for the apparently outstanding customer satisfaction performance.

An AT&T trouble-shooting team discovered that one of the critical factors explaining the paradox was the way in which Customer Satisfaction was being measured …

AT&T assembled a team to find out why there was no apparent connection between customer satisfaction (95%) and business performance (market share down 6%), and to fix it. As described by Kordupleski (2003), this team identified three core issues:

- The first was what they were doing with their raw data. Customer satisfaction was being measured on a 4-point scale: *Poor*, *Fair*, *Good* and *Excellent*. The *Good* and *Excellent* responses were being combined into a single *Satisfied Customer* category, giving rise to the 95% score for customer satisfaction. This was a major mistake. Of those who had rated them Excellent, almost all were very willing to repurchase from AT&T. In contrast, of customers who had rated them *Good*, some 40% were not very willing to buy again and were shopping for an alternative provider.
- The second mistake was to not benchmark their customer satisfaction scores against those of the competition. After all, business is a competition and customers have choices. If competitors do a better job at satisfying customers you will lose market share.
- The third and most important mistake was a failure to focus on Value as the ultimate metric, where Value was defined as the trade-off between people's satisfaction with the Quality of the product or service they were receiving balanced against their satisfaction with the Price paid. In simple words, did customers perceive the products and services received to be "worth what they paid"?  AT&T developed a value metric called CVA or Customer Value Added and deployed a process called Customer Value Management (CVM). It lead to a major turn-around in AT&T's fortunes (an increase of 7 points of market share in a year and a half as reported by the *Wall Street Journal*) and, subsequently, the fortunes of many enterprises world-wide who learned and adopted the approach . Fuller descriptions of the following very terse summary of CVM can be found in Kordupleski (2003) and Fisher (2013).

At the heart of the CVM toolset is the Customer Value Tree.  Let's take as an example the purchase of an automobile.  The overall concept of Value (*Worth What Paid For*) is modeled as having two principal drivers, Quality and Price, each of which can also be elaborated, as shown in Figure 1.

The Customer Value tree forms the basis of a survey of the market (both your customer and those of your competitors); *importantly, it is a survey of people who make the purchasing decisions*.  The Customer Value tree also forms the basis of the report of the results displayed in a simple to understand and use format. Respondents are asked to rate the performance of their supplier on **Automobile** Attributes (on a scale of 1 to 10, where 1 = *Poor* and 10 = *Excellent*). Then they are requested to provide an overall rating of the **Automobile**, together with the main reason for assigning this overall rating.  The rating process continues for the whole tree, up to an overall rating of **Worth What Paid For**.  At this point, it is useful to request higher-level ratings of business impact, such as **Willingness to repurchase** or **Willingness to recommend your company to someone else**.

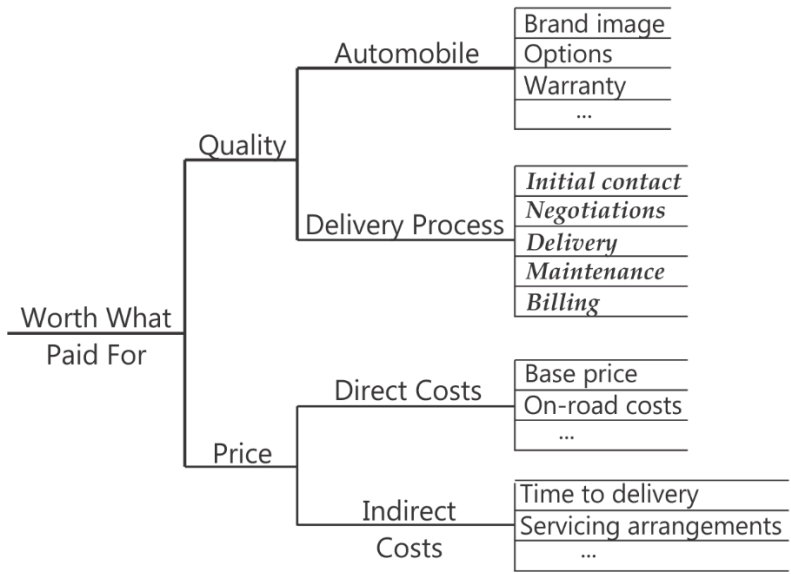

Figure 1. A prototypical Customer Value tree for buying and using an automobile. Value (Worth What Paid For) is represented as having two main Drivers, Quality and Price. Quality has as its Drivers the Product (in this case, an automobile) and *Delivery Process* – the sequence of experiences (*service sub-processes*) when the customer interacts with the supplier. Automobile, Direct Costs and Indirect Costs each have 6 – 7 Attributes determined from market focus groups. In some cases, Brand Image may be sufficiently important as to be elevated to the same status as Quality and Price as a Driver of overall Value.

Thus the overall response from a respondent takes the form of a tree-structured set of ratings.

A sample of such tree-structured data can then be analyzed by a fitting a sequence of hierarchical regression models: **Automobile** as a function of its Attributes, …, **Delivery Process** as a function of its constituent sub-processes, …, all the way up to **Value** as a function of **Quality** and **Price**. This modeling process yields two critical sets of information:

- the hierarchy of fitted models, which provide confirmation, or otherwise, that no important factor affecting the market's overall perception of Value has been omitted; and

- for each model, the relative rating of each explanatory variable and its impact weight.

The first endows this approach to perception surveys with a unique advantage over other approaches. And the second provides the basis for a very powerful management decision process to focus improvement efforts. We sketch a simple example.

Table 1 shows the top-level profile for your company and the average of your competitors. Overall, you are somewhat below par on Relative Value and Relative Satisfaction with Quality, and around par on Relative Satisfaction with Price.

| Driver | Impact weights (%) | Mean ratings (± 0.2) | | Relative rating (%) |
| --- | --- | --- | --- | --- |
| | | Our company | Competitors | |
| Quality | 51 | 7.4 | 7.7 | 96 |
| Price | 35 | 7.1 | 7.0 | 101 |
| **Value** | ($R^2$ = 81%) | 7.3 | 7.5 | CVA = 97 |

Table 1. Top-level table of impact weights and comparative ratings for Value and its main drivers. The Relative Value metric is known as CVA (Customer Value Added).

So, we look at two issues:

- how to connect the current rating of Value to business performance
- how to use the results to select and act on improvement priorities

Figure 2 is an example of a Loyalty curve. Recall that we had collected respondent data on **Willingness to recommend**. Conventionally, a rating of 8, 9 or 10 is regarded as in indicative of a respondent being very willing to recommend your product or service (although it could be defined as just a rating of 9 or 10). Figure 2 indicates that the current Value score of 7.3 corresponds to about 63% of your customers being very willing to recommend. To see an increase in this figure to, say, 80% will require an increase in the Value score to around 7.8.

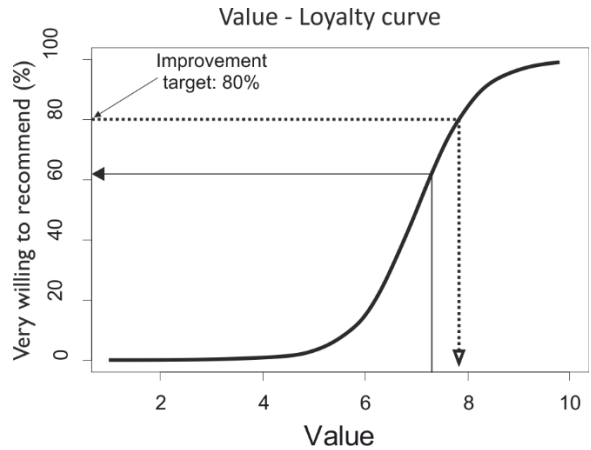

Figure 2. The current Value rating of 7.3 corresponds to just 63% of your customers being very willing to re-purchase from you. If you want this to be at least 80% within 12 months, you'll need to raise your Value score to about 7.8.

Where should improvements be focused? Suppose you could effect an improvement of 0.6 in the rating for Quality over the next 12 months (an analogous analysis for Price is also appropriate). Then, from Table 1, the predicted increase in Value would be simply $0.51 \times 0.6 \approx 0.31$. Table 2 shows the corresponding profile for Quality –

| Driver | Impact weights (%) | Mean ratings (± 0.2) | | Relative rating (%) |
| --- | --- | --- | --- | --- |
| | | Our company | Competitors | |
| Automobile | 39 | 7.8 | 7.5 | 104 |
| Del. process | 59 | 6.9 | 7.8 | 88 |
| **Quality** | ($R^2$ = 89%) | 7.4 | 7.7 | 96 |

Table 2. Profile of impact weights and comparative ratings for Quality and its main drivers.

– and drilling down further we get Table 3, the profile for the Delivery process and its sub-processes. It appears that there are serious problems for customers interacting with your finance department. Now you know where to focus attention.

| Driver | Impact weights (%) | Mean ratings (± 0.2) | | Relative rating (%) |
|---|---|---|---|---|
| | | Our company | Competitors | |
| *Initial contact* | … | | | |
| *…* | … | … | … | … |
| *Billing* | 40 | 6.1 | 7.5 | 81 |
| **Delivery process** | ($R^2$ = 86%) | 7.4 | 7.7 | 96 |

Table 3. Profile of impact weights and comparative ratings for Delivery process and its sub-processes.

The actions to take are:

- Develop a lower-level tree for Billing (Figure 3), and put in place some internal process metrics to track improvements. *Note that, in contrast with a Value survey which is focused on decision-makers, this survey is actually focused on actual users, a critical issue when we come to clarifying the characteristics of NPS.*

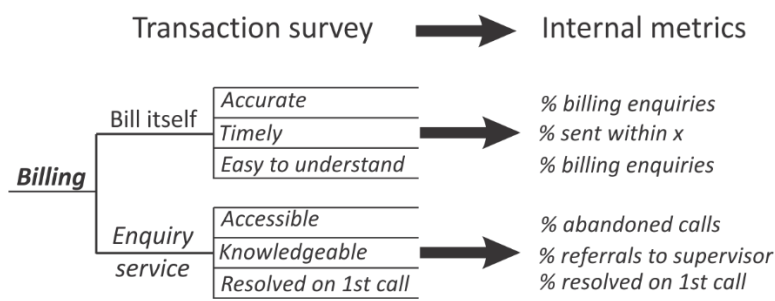

Figure 3. Billing tree for a transaction survey, together with internal metrics that can be tracked to monitor improvements.

- Carry out a Transaction survey, focused just on your own customers, to ascertain where the specific issues are with Billing.

- Make appropriate improvements, using the internal metrics to confirm stabilization of the Billing process and monitor it.

We are now in a position to capture the CVM continuous improvement process in a simple diagram (Figure 4).

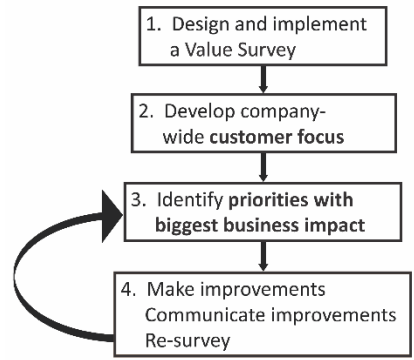

Figure 4. The CVM improvement cycle.

It is easy verified that this process meets the five criteria (a) – (e) outlined at the beginning of the Section (but see Section 4 for further discussion of the benchmarking issue). In summary, then, CVM is a robust, rigorous and proven process for gaining and sustaining a company's competitive position in the market. See Kordupleski (2003, 2018) for numerous applications. It is based on an ongoing cycle of continuous improvement, building and sustaining the relationship between the company and its market by monitoring and responding to current and emerging market needs and its competitive performance in relation to these needs.

## 2. Comparison with Net-Promoter Score

Net-Promoter Score (hereafter NPS) was introduced by Reichheld in a 2003 issue of *Harvard Business Review*, in an article entitled "The One Number You Need to Grow". The article led off with the breathless announcement:

> If growth is what you are after, you won't learn much from complex measurements of customer satisfaction or retention. You simply need to know what your customers tell their friends about you.

So, what is this remarkable number?  Reichheld (*op. cit.*) provided the following description:

> Asking a statistically valid sample of customers "How likely is it that you would recommend out company to a friend or colleague?" enables you to calculate your **net-promoter score**: the ratio [*sic*] of promoters to detractors.
>
> Based on their responses on a 0 to 10 rating scale, group your customers into "promoters" (9–10 rating – extremely likely to recommend), "passively satisfied" (7–8 rating), and "detractors" (0–6 rating – extremely unlikely to recommend).  Then subtract the percentage of detractors from the percentage of promoters.

And readers were advised of the very substantial benefits likely to flow:

> Many companies – striving for unprecedented growth by cultivating intensely loyal customers – invest lots of time and money measuring customer satisfaction.  But most of the yardsticks they use are complex, yield ambiguous results, and don't necessarily correlate to profits or growth.
>
> The good news is: you don't need expensive surveys and complex statistical models … By asking this one simple question you collect simple and timely data that correlate with growth.  You also get responses you can easily interpret and communicate. Your message to employees – "Get more promoters and fewer detractors" – becomes clear-cut, actionable, and motivating, especially when tied to incentives.

Abraham is famously said to have opined: "You can fool all the people some of the time and some of the people all the time, but you cannot fool all the people all the time."  Sadly, the first part of this dictum has proved correct.  Leaders of numerous enterprises that were using rigorous ongoing market research to gain and maintain competitive advantage were seduced by the twin prospects of massive savings and greatly improved business performance, cancelled their market research campaigns, and signed up to NPS with Reichheld's company.  Even 15 years later, NPS is exercising a malevolent influence on some of the world's largest corporations. For example, at the time of writing, Australia's four largest banks are literally in the dock (in the form of a [Royal Commission into Misconduct in the Banking, Superannuation and Financial Services Industry](#)), because of their treatment of their customers. And what is the principal metric banks have been using to boast about their performance with customers? NPS.

Much of Reichheld's article amounts to a simple statement of the obvious: the importance of customer loyalty – whether expressed as repeat purchasing of the same product or service, purchasing across the range, or recommendations to others – and how this "correlates to" higher-level business drivers such a market share and profitability. (It is interesting that this had to be explained to readers of *Harvard Business Review*.) However, there is some space devoted to debunking other approaches. For example:

> One of the main takeaways from our research is that companies can keep customer surveys simple. The most basic surveys employing the right questions – can allow companies to report timely data that are easy to act on. Too many of today's satisfaction survey process yield complex information that's months out of date by the time it reaches frontline management. Good luck to the branch manager who tries to help an employee interpret a score from a complex weighting algorithm based on feedback from anonymous customers, many of whom were surveyed before the employee got his current job.

Where do we find fault with NPS?

i. Many examples in Reichheld's article are of operational symptoms, without insight into root causes (systemic or otherwise). See for example the first main paragraph in the second column on page 4:

> Even the most sophisticated satisfaction measurement systems have serious flaws. I saw this first hand at one of the Big Three car manufacturers. The marketing executive at the company wanted to understand why, after the firm had spent millions of dollars on customer satisfaction surveys, satisfaction ratings for individual dealers did not relate very closely to dealer profits or growth. When I interviewed dealers, they agreed that customer satisfaction seemed a reasonable goal.  But they also pointed out that other factors were far more important to their profits and growth, such as keeping pressure on salespeople to close a high percentage of leads, filling showrooms with prospects through aggressive advertising, and charging customers the highest possible price for a car.

Indeed, as our introductory scenario suggested, an invitation to provide an NPS rating can be triggered by the most low-level of customer experiences.  Thus, an administrative assistant using an Enquiry service (*e.g.* Figure 3) who was frustrated by being placed on hold for a lengthy period might vent that frustration by a very low NPS rating when asked to say on the line at the end of the call and answer a question.   Does a company really want to have its whole performance judged by an instant of minor irritation?  The contrast with a Customer Value survey is critically important:

- In a Customer Value process, the person being surveyed is a decision-maker. The respondent is asked to rate all aspects of the customer experience before assigning a rating for overall Value. Only at that point is the respondent asked about *Willingness to Recommend*.
- In an NPS process, the person being surveyed is typically a user.  Rather than being led through the whole customer experience before being asked about *Willingness to Recommend*, the respondent is being asked to make a decision based on one particular interaction with the company, no matter how trivial.

ii. Table 4 provides an assessment of NPS against the five criteria in Section 2:

| | | |
|---|---|---|
| a. | A statistically sound method that ensures that no important attributes of the product or services have been omitted from the survey. | Totally ignored. The current routine collection of NPS scores for operational activities (*cf.* the introductory example in Section 2) is purely observational, with little or no understanding of demographic factors let alone sampling biases. |
| b. | A means of linking survey results to higher-level business drivers. | Assertion of "correlation with growth and profitability". |
| c. | Actionable survey results, including the ability to drill down. | No. |
| d. | A means of identifying where to focus improvement priorities so as to have the greatest beneficial impact on the business bottom line. | No defensible method. |
| e. | Meaningful benchmarking metrics. | Nothing below level of NPS. Also, there are no agreed standards about how NPS should be aggregated in a company to produce an all-of-company metric. |

Table 4. Evaluation of NPS against desiderata for satisfaction surveys.

iii. Of course, there are plenty of poorly constructed market research surveys. However there are good ones as well, and any comparison should be with what is best, not what is worst.

iv. Correlation doesn't equate to causation.

v. A company using NPS is basically not distinguishing those of its customers rating them 0 from those rating them 6, and is totally ignoring 'Passives', who are rating them 7 or 8. This is a recipe for losing market share. Those rating them 0 – 3 or 4 are almost certainly going to leave. However, with more information about customer needs, those rating them 5 or 6 could possibly be converted to Passives, and those currently categorized as Passives could be boosted to Promoter status.

vi. As is well-known from the AT&T work 30 years before, what is critical is the perception of Relative Value: there are many ways of measuring 'customer satisfaction', some OK, many others not OK, but *none of which Reichheld mentions.*

vii. In the case of AT&T, some of the 9 Business units were operating Business to Business, a few were purely just Business to Consumer, but most were both. The CVA metric was predictive of market share for all industries, markets and countries. For Business to Business, NPS is even weaker because the decision maker is hardly ever surveyed. AT&T standards required CVA to be calculated from decision-maker surveys of the market that were random in time and sample, and not triggered by a recent event.

viii. Many executives use NPS as a way to pass on the responsibility for satisfying customer needs to a coal-face employee so that the executives can be free to focus on satisfying the bottom line and the company's investors. However no one employee can fulfil a customer's total and real needs, especially in the absence of guidance about where the real problems lie. It is the senior leadership who need to be able to see the whole picture, rather than just a single, often self-serving metric. Only by carefully evaluating what can be done to improve products, services and costs can a company improve the quality of life for its customers.

ix. As is described by Kordupleski (2018), the realization by AT&T in 1986 that they needed to focus on Value, and Relative Value resulted from extensive statistical exploration of a very large market research data base, with considerable time being spent seeking correlations and cross-correlations to high-level business performance indicators such as Market Share and Return on Invested Capital (ROIC). Thus, for example, CVA emerged as a *lead indicator* of Market Share, not simply as a metric correlated to Market Share. CVA, a customer satisfaction-derived marketing metric could, using a metaphor, "predict

x. the weather." What is even more important, a firm could use the data and information to "improve the weather."

x. It is important to note that indicators such as Market Share and ROIC *relate to the entire market*, not just to one's own customers, whereas Loyalty, which is fine as far as it goes, still relates just to one's own customers. This brings to mind a series of annual Quality conferences in Australia where marketing for the next meeting was focused primarily on attendees at the previous meeting. (The principal organizer of the conferences couldn't understand how, if every conference had nearly 90% satisfaction rating, attendance had fallen from around 1200 at an initial meeting to around 450 ten years later. $1200 \times (0.9)^9 \approx 464$.)

We note that NPS has not been entirely overlooked by statisticians as an object of study. Jeske, Callanan and Guo (2011) studied NPS and the claims made for it by Reichheld, and made the following insightful observations: :

> … the hope is that movements in NPS are positively correlated with revenue growth for the company. While Reichheld's research presented some evidence of that, other findings are not as corroborative … . Regardless of whether there is a predictive relationship between NPS and revenue growth, implementing policies and programs within a company that improve NPS is an intuitively sensible thing to do … . A difficult and important question, however, is how to identify key drivers of NPS. *Calculating* NPS alone does not do this.

They then proceed to show how to gain insight into what these key drivers might be by applying statistical modeling to an existing customer survey. This analysis is fine, as far as it goes. However, it was conducted (seemingly) in the absence of knowledge of the CVM process and thus of the knowledge gained from a wealth of case studies identifying Value, and Relative Value, as the crucial lead indicators of high-level business drivers (market share, ROIC, …) as spelt out by Kordupleski (2003, 2018). Further, amongst the valuable tools associated with CVM is the concept of a Value Map (Figure 5), which a company can used for strategic positioning. *Inter alia*, the Value Map gives meaning to the concept of a Value Proposition. First a company selects which market (Economy, Average or Premium) it is targeting on the Value Map. Then it decides whether it will gain market superiority by being superior on Quality and at par on Price, or by being superior on Price and at par on Quality, or both.

In fact, Kordupleski (2018) is largely a published version of an unpublished article prepared in 1989. To quote from its Abstract:

> The following article, "The Right Choice – What Does It Mean" by R.E. Kordupleski and W. C. Vogel, Jr. is a 1989 paper that reported on some of the most significant findings in the early days of customer value measurement and management. It is based on one of the largest empirical data bases available at that time. AT&T was doing over 60,000 customer surveys per month. Three years of monthly findings were analyzed by some of the best researchers and scientists in the 300k employee and $85 billion annual revenue company. The paper presented empirical evidence of the power of the consumer's perception of customer value, its impact on market share, growth and customer loyalty and ultimately its impact on shareholder value and employee value. The paper was never published, but it was released throughout AT&T and also to AT&T's strategic business partners. Its content was presented and discussed at national conferences hosted by the American Marketing Association and the US Conference Board. Its findings have stood the test of time.

Clearly, there need to be more bridges built between the statistical and market research literatures!

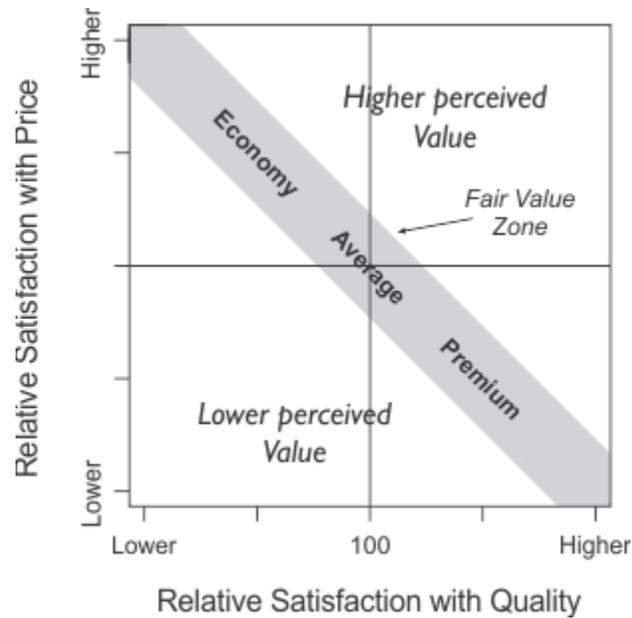

Figure 5, Value Map. The scores for Relative Quality and Relative Price are plotted against those of the competitors. The Fair Value zone relates to which market sector you are targeting. Adapted from Fisher (2013, Figure 5.6), based on the concept described by Kordupleski (2003); see also Kordupleski (2018).

### 3. Net-Promoter Score – a false god

Customer satisfaction research needs to answer two simple questions

- How are we going?

and

- What should we do?

NPS does only a modest job answering the first but contributes nothing towards answering the second.

For Customer Value Management,

A. Extensive statistical analysis from a large amount of market research data has established that Relative Value is a lead indicator for competitive business outcomes such as Market Share and ROIC.

B. Relative Value is derived from data collected from decision-makers (people making the purchasing decisions), and necessarily involves acquiring competitive data as well as data from your own company. It is calibrated by competitive information.

C. The decision-maker's rating of Value is arrived at only after consideration of the entire customer experience, *involving both Quality and Price*. The initial failure by AT&T to obtain a connection between customer satisfaction and business performance was due in part to overlooking satisfaction with Price.

D. There is a clearly defined and proven process for identifying improvement priorities likely to have the greatest beneficial impact on the business.

For NPS,

a. There appears to be little hard evidence that NPS is a lead indicator of business outcomes; indeed the AT&T experience suggests that there may be no connection.

b. NPS focuses only on the user who, in many cases, will not be the person making the purchasing decision. It is uncalibrated by competitive information.

c. It often derives from a single customer experience with unknown influence on a decision-maker's overall perception of Value.

d. There is no sound approach to selection of improvement priorities.

There are well-established 'best practice' approaches to creating and delivering superior value to customers and so gaining and sustaining market share, and there are associated metrics. NPS is not such a metric: it is a false god.